\documentclass[%
reprint,
superscriptaddress,
onecolumn,
amsmath,
amssymb,
aps,
prl,
nolongbibliography
]{revtex4-2}

\usepackage[T1]{fontenc}
\usepackage[utf8]{inputenc}

\usepackage{times} 
\usepackage[unicode=true,breaklinks=true,colorlinks=true]{hyperref}
\hypersetup{citecolor=blue,urlcolor=blue}
\usepackage{graphicx}
\usepackage{bm}
\usepackage{bbold}
\usepackage{braket}
\usepackage[dvipsnames]{xcolor}

\usepackage{mathtools}
\usepackage{tensor}

\newcommand\norm[1]{\left\lVert#1\right\rVert}

\usepackage[normalem]{ulem}

\begin{document}
\title{Notes on Thermodynamic Principle for Quantum Metrology}
\author{Yaoming Chu}
\affiliation{School of Physics, Hubei Key Laboratory of Gravitation and Quantum Physics, Institute for Quantum Science and Engineering, International Joint Laboratory on Quantum Sensing and Quantum Metrology,  Huazhong University of Science and Technology, Wuhan 430074, China}
\affiliation{Wuhan National High Magnetic Field Center, Wuhan National Laboratory for Optoelectronics, Huazhong University of Science and Technology, Wuhan 430074, China}
\author{Jianming Cai}
\email{jianmingcai@hust.edu.cn}
\affiliation{School of Physics, Hubei Key Laboratory of Gravitation and Quantum Physics, Institute for Quantum Science and Engineering, International Joint Laboratory on Quantum Sensing and Quantum Metrology,  Huazhong University of Science and Technology, Wuhan 430074, China}
\affiliation{Wuhan National High Magnetic Field Center, Wuhan National Laboratory for Optoelectronics, Huazhong University of Science and Technology, Wuhan 430074, China}

\begin{abstract}
Recently, we find a physical limit on energy consumption of quantum metrology, and demonstrate that it essentially arises from the erasure of quantum Fisher information (QFI) which determines the best achievable measurement precision. Here, we provide more details in order to further elaborate the essence of this principle.  
\end{abstract}

\maketitle

\subsection{The main result: Thermodynamic principle for quantum metrology} 

We reiterate the main result of our Letter by quoting: {\it “As our main result, we find the physical limit on energy consumption of quantum metrology, and demonstrate that it essentially arises from the erasure of quantum Fisher information (QFI) which determines the best achievable measurement precision”} \cite{Chu2022}. 

\begin{figure}[!h]
\centering
\includegraphics[width=12cm]{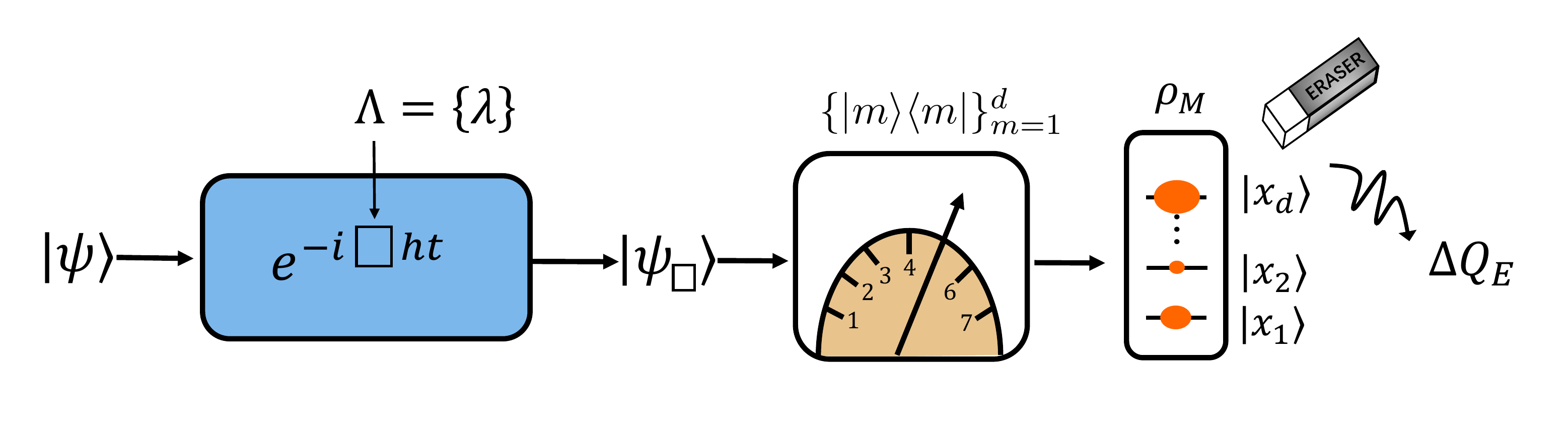}
\caption{Schematic representation of a quantum metrology machine and its overall thermodynamic performance in terms of the average heat dissipation $\Delta Q_E=\langle \Delta Q_E^\lambda\rangle_{\lambda}$, which originates from recovering the memory's final state after the measurement (i.e. characterized by the ensemble $\rho_M$) to its standard state.}
\label{fig:protocol}
\end{figure}

The principle is explicitly presented in Eq.\,(8) of our Letter \cite{Chu2022}, where we consider the traditional pure-state quantum metrological scenario (see Fig.\ref{fig:protocol}), namely the interrogation Hamiltonian is in the form of $H_\lambda=\lambda h$ and the $\lambda$- parametrized state of the system is prepared via $|\psi_\lambda\rangle=e^{-i\lambda h t}|\psi\rangle$ with $t$ the interrogation time. After a general measurement in the basis of $\{|m\rangle\langle m|\}_{m=1}^d$ ($d$ is the system dimension) to extract the information on the parameter $\lambda$, the following recovery of the memory to its original state would result in heat dissipation into the environment. Given an environment of temperature $T$, we find that the average heat dissipation $\Delta Q_E$ over the unknown parameter $\lambda$ is lower bounded by the QFI $F_Q$ of the state $|\psi_\lambda\rangle$ about the parameter  (assuming $\norm{h}=1$) as
\begin{equation}
\label{eq:Q-Fq-average}
 \Delta Q_E \geq \log (2) k_B T t^{-2} F_Q.
\end{equation}
The validity of the above inequality is proved in the section of “{\it Average heat dissipation in quantum metrology}” (cf. Supplementary Materials \cite{Chu2022}). Moreover, the principle is illustrated using the example of quantum Rabi model and is extended to the scenario of multiqubit quantum metrology by introducing the weighted QFI, which highlights an efficient way to investigate energy efficiency of multiqubit quantum states. Below we clarify the three main ingredients that can help to understand the principle in Eq.\eqref{eq:Q-Fq-average}: (1) The definition of average heat dissipation; (2) The bound of average heat dissipation given by the Shannon entropy; (3) The bound of the Shannon entropy by the QFI.

\subsubsection{The definition of average heat dissipation}

In Landauer's principle, the energy consumption for erasing a bit (namely $k_B T$) is quantified by the average heat dissipation of {\it a universal protocol erasing a bit state randomly chosen from both state 0 and 1} \cite{Bennett2003,Reeb2014}. Similarly, the established thermodynamic principle for quantum metrology is formulated for a quantum metrology machine (with {\it a fixed measurement apparatus}, see Fig.\ref{fig:protocol}) in terms of the average heat dissipation over metrological runs for all possible values of the unknown parameter $\lambda$, namely 
\begin{equation}
\label{Eq:Average_Q}
\Delta Q_E=\langle \Delta Q_E^{\lambda}\rangle_{\lambda},    
\end{equation}
where $\Delta Q_E^{\lambda}$ represents the heat dissipation for the parametrized state $\vert\psi_\lambda\rangle$. For a given measurement protocol $\{|m\rangle\langle m|\}_{m=1}^d$ ($d$ is the dimension of the system), i.e. 
\begin{equation}
  \label{eq:M}
  \mathcal{M}:|\psi_\lambda\rangle\langle\psi_\lambda| \otimes |x\rangle\langle x| \to \sum_m p_m^\lambda |m\rangle\langle m|\otimes |x_m\rangle\langle x_m|,
\end{equation}
where $p_m^\lambda=|\langle m|\psi_\lambda\rangle|^2$, and $\{|x_m\rangle\}$ represents the internal structure of a memory, interacting with the system and storing the measurement outcomes. A unitary operation conditional on the memory's state brings the system back to the initial state without heat dissipation into environment because of its reversibility \cite{Peres1985}. In order to realize a closed metrological cycle, the memory in the mixed state $\rho_M^\lambda=\sum_m p_m^\lambda|x_m\rangle\langle x_m|$ must be recovered to its original standard state $|x\rangle$, resulting in a heat dissipation $\Delta Q_E^\lambda$ into the environment. %

\subsubsection{The bound of average heat dissipation given by the Shannon entropy}

Similar to Landauer's principle \cite{Bennett2003,Reeb2014}, we consider the average heat dissipation (i.e. $\Delta Q_E$) over all possible values of the unknown parameter $\lambda$ for the given measurement basis $\{|m\rangle\langle m|\}_{m=1}^d$, and characterize the overall thermodynamic performance of a quantum metrology machine. In such a framework, the memory's state after the measurement protocol can be expressed as
\begin{equation}
    \rho_M=\langle \rho_M^\lambda\rangle=\sum_{m=1}^dp_m |x_m\rangle\langle x_m|.
    \label{Eq:rho_M}
\end{equation}
The probability distribution $\{p_m\}_{m=1}^d$ is given by
\begin{equation}
    p_m = \frac{1}{\mathcal{N}}\int p_m^\lambda d\lambda = \frac{1}{\mathcal{N}}\int \langle m|\psi_\lambda\rangle\langle\psi_\lambda|m\rangle d\lambda = \langle m|\rho_s|m\rangle,
\end{equation}
where $\mathcal{N}$ is the normalization factor to keep $\sum_{m}p_m=1$, and
\begin{equation}
\rho_s=\frac{1}{\mathcal{N}}\int|\psi_\lambda\rangle\langle\psi_\lambda|d\lambda, 
\label{eq:rho_s}
\end{equation}
related to the parametrized-state ensemble $\{|\psi_\lambda\rangle\langle\psi_\lambda|\}$. According to Landauer's principle, recovering the memory to its standard state inevitably dissipates a certain amount of heat lower bounded by the entropy $\mathcal{S}=-\operatorname{tr}(\rho_M\log\rho_M)$ \cite{Bennett2003,Reeb2014}, namely
\begin{equation}
\label{Eq:Average_Q_2}
    \Delta Q_E \geq k_B T\mathcal{S}= -k_B T\sum_{m=1}^d p_m \log(p_m).
\end{equation}
We emphasize that the above averaging procedure to characterize the quantum metrology machine is physically reasonable and meaningful, which results from the lacking of a priori information about the unknown parameter to be estimated in quantum metrology.

\subsubsection{The bound of the Shannon entropy by the QFI}

For the traditional quantum metrological scenario [i.e. Eq.\,(5) in our Letter \cite{Chu2022} where the interrogation Hamiltonian $H_\lambda=\lambda h$], we prove that (cf. Supplementary Materials \cite{Chu2022})
\begin{equation}
\mathcal{S}\geq S(\rho_s)\geq \log (2) t^{-2} F_Q.
\label{Eq:averageS}
\end{equation}
Here, $\mathcal{S}=-\operatorname{tr}(\rho_M\log\rho_M)=-\sum_{m=1}^d p_m \log(p_m)$ denotes the entropy of the memory after the measurement, $S(\rho_s)=-\operatorname{tr}(\rho_s\log\rho_s)$ [see Eq.\eqref{eq:rho_s} for the definition of $\rho_s$] and $F_Q$ represents the QFI of the parametrized quantum state $|\psi_\lambda\rangle=e^{-i\lambda h t}|\psi\rangle$. Note that $F_Q$ is independent of the value of the parameter $\lambda$ and can be interpreted as the QFI associated with the state preparation,
\begin{equation}
\mathcal{P}:\,\,|\psi\rangle \to |\psi_\lambda\rangle=e^{-i\lambda h t}|\psi\rangle.
\end{equation}
Based on Eq.\,\eqref{Eq:Average_Q_2} [which is essentially Landauer's principle] and Eq.\,\eqref{Eq:averageS}, we can establish the following thermodynamic principle [i.e. Eq.\,(8) in our Letter],
\begin{equation}
\Delta Q_E\geq k_B T \log (2) t^{-2} F_Q,
\label{Eq:averageQ}
\end{equation}
which implies that the thermodynamic cost of a quantum metrology machine is lower bounded by the QFI.
\subsubsection{Another single-qubit example}

\begin{figure}[!b]
\centering
\includegraphics[width=75mm]{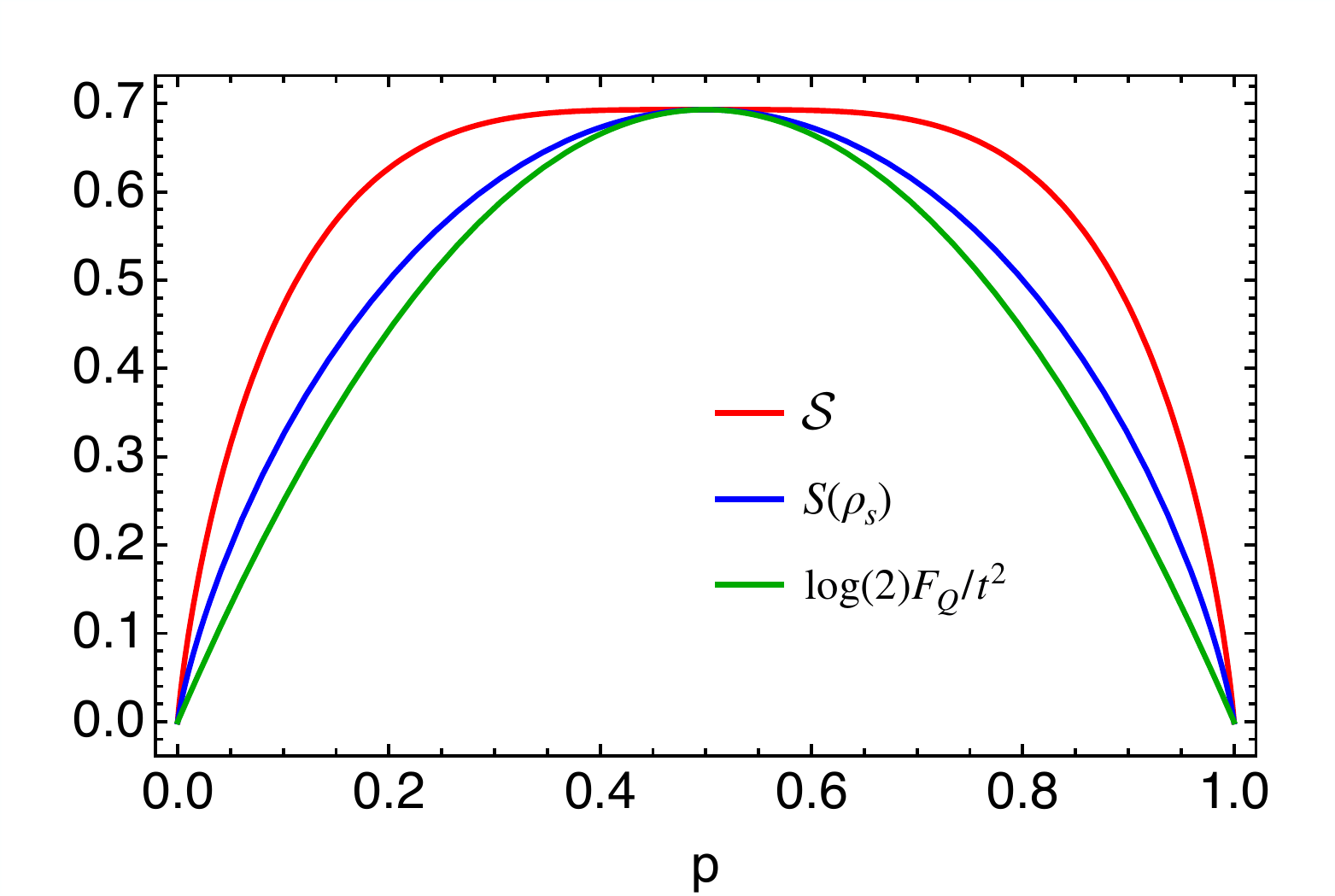}
\caption{Illustration of Eq.\,\eqref{Eq:averageS} $\mathcal{S}\geq S(\rho_s)\geq \log(2)F_Q/t^2$ (cf. Eq.(7) in our Letter \cite{Chu2022}) with a single-qubit example. The result demonstrates that the Shannon entropy of the measurement outcomes $\mathcal{S}$ is well bounded by the von Neumann entropy $S(\rho_s)$ and the quantum Fisher information as $\log(2)t^{-2}F_Q$.}
\label{fig2}
\end{figure}

Note that Eq.\,\eqref{Eq:averageS} is proved for a general measurement protocol, including optimal and non-optimal measurements. In our Letter \cite{Chu2022}, we have demonstrated the application of the established thermodynamic principle in quantum Rabi model. Here, we further illustrate the principle by considering another example, namely a non-optimal measurement protocol in a single-qubit system. The qubit is prepared into the following parametrized state
\begin{equation}
|\psi_\lambda\rangle=\exp [-i(\lambda \sigma_z/2)  t](\sqrt{p}|0\rangle+\sqrt{1-p}|1\rangle),
\label{eq:psi_lambda}
\end{equation}
with the interrogation Hamiltonian 
\begin{equation}
H_{\lambda}=\lambda h=\lambda (\sigma_z/2).
\end{equation}
The parametrized state is measured in the basis of $\{|\pm_p\rangle\langle \pm_p|\}$ with 
\begin{eqnarray}
|+_p\rangle&=&\sqrt{p}|0\rangle+\sqrt{1-p}|1\rangle,\\
|-_p\rangle&=&\sqrt{1-p}|0\rangle-\sqrt{p}|1\rangle.
\end{eqnarray}
The corresponding probabilities can be obtained as
\begin{equation}
\begin{aligned}
p_{+}&=2p^2-2p+1+2p(1-p)\cos(\lambda t),\\
 p_{-}&=2p(1-p)-2p(1-p)\cos(\lambda t).
 \end{aligned}
\end{equation}
Hence, the average probabilities in the state $|+_p\rangle$ and $|-_p\rangle$ over all the possible values of $\lambda$ are given by 
\begin{equation}
 \tilde{p}_{+}=\langle p_+\rangle_\lambda=2p^2-2p+1, \quad \tilde{p}_-=\langle p_-\rangle_\lambda=2p(1-p),
\end{equation}
resulting in the Shannon entropy of the measurement outcomes as 
\begin{equation}
\mathcal{S}=-\tilde{p}_+\log(\tilde{p}_+)-\tilde{p}_{-}\log(\tilde{p}_{-}).
\end{equation}
The von Neumann entropy $\rho_s$ [see Eq.\,\eqref{eq:rho_s}] is written as
\begin{equation}
S(\rho_s)=-p\log(p)-(1-p)\log(1-p).
\end{equation}
And the QFI corresponding to the parametrized state [see Eq.\,\eqref{eq:psi_lambda}] is 
\begin{equation}
F_Q=4p(1-p).
\end{equation}
Thus, it is easy to verify that Eq.\,\eqref{Eq:averageS} holds. Furthermore, it can be seen from Fig.\ref{fig2} that the Shannon entropy (thereby the heat dissipation according to Landauer's principle) and the QFI shows strongly correlated behavior in this example.

\subsection{Relation between the Shannon entropy and the QFI for SLD measurement}

\subsubsection{A generalized inequality}

Apart from the main result of thermodynamic principle for quantum metrology based on the {\it average heat dissipation over the unknown parameter} (as we elaborate in the above section), in Ref.\,\cite{Chu2022}, we also present a relation between the Shannon entropy associated with the SLD measurement and the QFI for {\it a fixed value of the unknown parameter} in a pure-state metrology, cf. Eq.(3) in Ref.\,\cite{Chu2022}. We would like to point out that such a relation can be replaced by the following more generalized inequality with $2 \norm{h_\lambda} \rightarrow \norm{L_\lambda}$, namely
\begin{equation}
\label{Eq:entropy_QFI}
\mathcal{S}\geq 4 \log(2) \norm{L_\lambda}^{-2} F_Q[\rho_\lambda],
\end{equation}
which holds for both pure and mixed states. Here, $\mathcal{S}$ is the Shannon entropy associated with the SLD measurement, $L_\lambda$ is the SLD operator, $F_Q[\rho_\lambda]$ is the QFI of a general mixed state dependent on the parameter $\lambda$. In more detail, the Shannon entropy of the measurement outcomes can be expressed as 
\begin{equation}
\mathcal{S}=-\sum_\ell p_\ell^\lambda \log (p_\ell^\lambda), 
\end{equation}
where $p_\ell^\lambda=\operatorname{tr}(|\ell_\lambda\rangle\langle\ell_\lambda|\rho_\lambda)$ with $|\ell_\lambda\rangle$ the eigenvector of $L_\lambda$ associated with the eigenvalue $\ell_\lambda$. The QFI of $\rho_\lambda$ with respect to the unknown parameter $\lambda$ can be related to the variance of $L_\lambda$ [Note that $\operatorname{tr}(L_\lambda \rho_\lambda)=0$] as 
\begin{equation}
  \label{eq:pure-QFI}
  F_Q[\rho_\lambda]=\operatorname{tr}(L_\lambda^2 \rho_\lambda)=\text{Var}[L_\lambda,\rho_\lambda]=\frac{1}{2}\sum_{\ell,\ell^\prime}(\ell_\lambda-\ell^\prime_\lambda)^2 p_\ell^\lambda p_{\ell^\prime}^\lambda.
\end{equation}
Hence, we have
\begin{equation}
2 F_Q[\rho_\lambda]/\norm{L_\lambda}^2\leq \sum_{\ell\neq \ell^\prime}p_\ell^\lambda p_{\ell^\prime}^\lambda=1-\sum_\ell (p_\ell^\lambda)^2.
\end{equation}
Based on Lemma 1 in Supplementary Material for our Letter \cite{Chu2022}, namely
\begin{equation}
  \label{eq:subsidiary}
  -\sum_\ell p_\ell^\lambda \log (p_\ell^\lambda) +2 \log 2\sum_\ell (p_\ell^\lambda)^2\geq 2\log 2,
\end{equation}
we obtain the generalized inequality in Eq.\,\eqref{Eq:entropy_QFI}, which represents a general version of Eq.(3) in our Letter \cite{Chu2022}.
As an example, we consider the following specific example as
\begin{equation}
\rho_\lambda=\frac{1}{2}\begin{pmatrix}
\lambda^2 & \lambda\\
\lambda & 2-\lambda^2
\end{pmatrix}.
\label{Eq:example}
\end{equation}
\begin{figure}[!b]
\centering
\includegraphics[width=70mm]{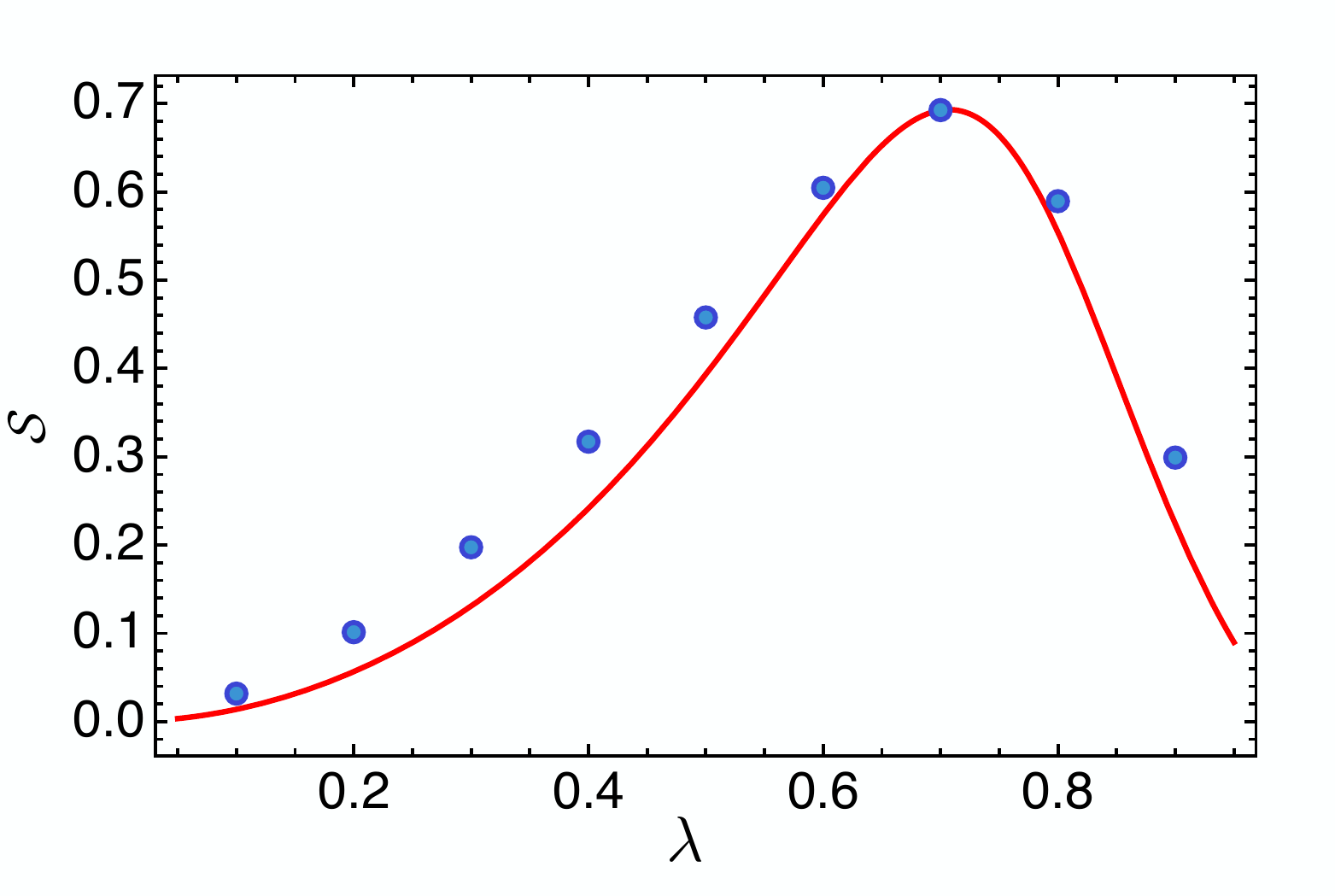}
\caption{The Shannon entropy $\mathcal{S}$ (blue dot) and its bound given by the QFI (red curve), cf. Eq.\,\eqref{Eq:entropy_QFI}, for the example of Eq.\eqref{Eq:example}.}
\label{fig}
\end{figure}
We numerically calculate the Shannon entropy $\mathcal{S}$, the semi-norm of the SLD operator $\norm{L_\lambda}$, and the QFI $F_Q[\rho_{\lambda}]$. The result in Fig.\ref{fig} shows the relation between the Shannon entropy $\mathcal{S}$ and the  bound given by the QFI [cf. Eq.\,\eqref{Eq:entropy_QFI}].

We remark that the generalized inequality in Eq.\,\eqref{Eq:entropy_QFI} can achieve the equality for two special cases: (i) $\rho_\lambda$ is pure (where $\norm{L_\lambda}=2 \sqrt{F_Q}$) \cite{Dooley2022}; (ii) $\rho_\lambda=p_1 |\phi_1(\lambda)\rangle\langle \phi_1(\lambda)|+p_2 |\phi_2(\lambda)\rangle\langle \phi_2(\lambda)|$ of a two-level system with $p_{1,2}$ independent of $\lambda$.
In these cases, both the Shannon entropy and the QFI bound in  Eq.\,\eqref{Eq:entropy_QFI} equal $\log(2)$. Note that $\norm{L_\lambda}\leq 2\norm{h_\lambda}$ for pure state (by using $L_\lambda=2i[\rho_\lambda,h_\lambda]$) in Eq.\,\eqref{Eq:entropy_QFI} and
gives rise to Eq.(3) in our Letter \cite{Chu2022}, which shall be replaced by Eq.\,\eqref{Eq:entropy_QFI} in order to provide meaningful connections between the Shannon entropy associated with the SLD measurement and the QFI, see an explicit example in the following section. We would like to stress that the main result of thermodynamic principle for quantum metrology is based on the {\it average heat dissipation over the unknown parameter}, while the relation between the Shannon entropy associated with the SLD measurement and the QFI is for {\it a fixed value of the unknown parameter}. The validity and applicability of the thermodynamic principle for quantum metrology established in our Letter does not rely on this relation.
%
%
\subsubsection{An example}

We remark that the generalized inequality in Eq.\,\eqref{Eq:entropy_QFI} may provide us interesting insights into the connections between the Shannon entropy and the QFI. Below we present an explicit example to illustrate this point. We consider a mixed state in the form of $\rho_\lambda=p_1(\lambda) |\phi_1(\lambda)\rangle\langle \phi_1(\lambda)|+p_2(\lambda) |\phi_2(\lambda)\rangle\langle \phi_2(\lambda)|$ in single-qubit system. In such a case, the SLD is given by \cite{Braunstein1994} 
 \begin{equation}
L_\lambda=\sum_{i}\frac{\partial_\lambda p_i}{p_i} |\phi_i\rangle\langle\phi_i|+(\Omega |\phi_1\rangle\langle \phi_2|+h.c.),
\end{equation}
with 
\begin{equation}
\Omega=2p_1\langle\partial_\lambda\phi_1|\phi_2\rangle+2p_2\langle\phi_1|\partial_\lambda\phi_2\rangle.   
\end{equation}
We note that the QFI can be divided as $F_Q=F_c+F_{nc}$ with the classical part $F_c=\sum_{i}(\partial_\lambda p_i)^2/p_i$ and the non-classical part $F_{nc}=|\Omega|^2$ \cite{Zanardi2007}. Thus, according to Eq.\,\eqref{Eq:entropy_QFI} we can further obtain 
\begin{equation}
\mathcal{S}\geq \log(2) \frac{F_Q}{F_Q+\left(\displaystyle \frac{1}{4p_1p_2}-1\right)F_{c}},
\label{Eq:S_QFI_TLS}
\end{equation}
the right-hand side of which is related to the ratio $\alpha \equiv F_{c}/F_Q$. Eq.\,\eqref{Eq:S_QFI_TLS} demonstrates that the Shannon entropy associated with the SLD measurement for quantum metrology with mixed states is connected not only to the total QFI but also with its classical and non-classical components. 
\subsection{Some remarks: The role of a priori information}

Firstly, we would like to remark that the generalized inequality in Eq.\,\eqref{Eq:entropy_QFI} pertains to the SLD measurement. For non-SLD measurement, the inequality may not be satisfied. In fact, based on the quantum version of Landauer's principle \cite{Reeb2014}, one can obtain that the heat dissipation of $|\psi_\lambda\rangle$ for a general measurement basis $\{|m\rangle\langle m|\}_{m=1}^d$ is lower bounded by
\begin{equation}
\Delta Q_E^\lambda\geq k_B T \mathcal{S}^\lambda\equiv -k_B T\operatorname{tr}(\rho_M^\lambda\log\rho_M^\lambda)=-k_B T\sum_{m=1}^d p_m^\lambda \log(p_m^\lambda).
\end{equation}
The right-hand side of the above inequality for pure-state metrology can be made arbitrarily small (i.e. $\mathcal{S}^\lambda\to 0$) by choosing a non-SLD optimal measurement basis as $\{\Pi_q=|q\rangle\langle q|,\Pi_{\bar{q}}=|\bar{q}\rangle\langle \bar{q}|\}_{q\to 1}$, where $|q\rangle=\sqrt{q}|\psi_{\lambda_0}\rangle+\sqrt{1-q}|\psi_{\lambda_0}^\perp\rangle$,  $|\bar{q}\rangle=\sqrt{1-q}|\psi_{\lambda_0}\rangle-\sqrt{q}|\psi_{\lambda_0}^\perp\rangle$ and $\lambda_0\to\lambda$ \cite{Dooley2022}. Note that the corresponding Fisher information of the measurement outcomes about the unknown parameter is maximized (i.e. equal to the QFI, which is non-zero) in this scenario. However, the entropy calculated in this example assumes a priori information about the parameter, such that the specific measurement basis can be chosen. We do not adopt this assumption for the formulation of a thermodynamic principle for quantum metrology in our Letter \cite{Chu2022}, as the role of a priori information about the unknown parameter needs to be properly taken into account. In general, $\Delta Q_E^\lambda$ would depend on how much a priori information about the unknown parameter is available. 

Secondly, we circumvent this difficulty by assuming no a priori information on the unknown parameter to be estimated. In our Letter \cite{Chu2022}, we establish the thermodynamic principle for quantum metrology based on the average heat dissipation over metrological runs for all possible values of the unknown  parameter $\lambda$, namely $\Delta Q_E=\langle \Delta Q_E^{\lambda}\rangle_{\lambda}$, where $Q_E^{\lambda}$ represents the heat dissipation for the parametrized state $\vert\psi_\lambda\rangle$. We stress that such a formulation, deeply connected to the erasure of the QFI (cf. Ref.\,\cite{Chu2022}, a "many-to-one" map $\mathcal{E}: |\psi_\lambda\rangle \to |\psi\rangle$ for all $\lambda$), is physically reasonable and justified, similar to Landauer's principle where the erasure of information represents $\mathcal{E}_L: 0/1 \to 0$ \cite{Bennett1982,Reeb2014}. 
Thirdly, we shall note that the principle [Eq.\eqref{eq:Q-Fq-average}, i.e. Eq.(8) in our Letter \cite{Chu2022}] is proved for a general measurement protocol (not limited to SLD measurement), and is applicable not only for optimal measurement but also for non-optimal measurement. As an example, we consider the single-qubit example \cite{Dooley2022}, i.e
\begin{equation}
 |\psi_\lambda\rangle=e^{-i\lambda\sigma_z/2}(|0\rangle+|1\rangle)/\sqrt{2}.
\end{equation}
The probability distribution under a fixed optimal measurement protocol (which achieves the maximal Fisher information) $\{|\pm\rangle\langle \pm|\}$ with $|\pm\rangle=(|0\rangle\pm e^{i\phi}|1\rangle)/\sqrt{2}$ is given by
\begin{equation}
    p_\pm^\lambda=|\langle \pm|\psi_\lambda\rangle|^2=\frac{1}{2}\left[1\pm\cos(\phi-\lambda)\right].
\end{equation}
In this case, $\langle p_\pm^\lambda\rangle=1/2$ and thus the Shannon entropy related to the measurement outcomes of such a quantum metrology machine is given by $\mathcal{S}=\log(2)$, which equals the right-hand side of Eq.\,\eqref{Eq:averageS}, i.e. $\log(2) F_Q=\log(2)$, and thus does not lead to any violation of the principle.
\subsection{Summary $\&$ Outlook}

To summarize, the main results about the thermodynamic principle of quantum metrology [i.e.\,Eqs.\,(7-11) in our Letter \cite{Chu2022}], which are formulated based on the average heat dissipation over the unknown parameter assuming no a priori information, characterize the overall thermodynamic performance of a quantum metrology machine and hold for general measurement protocols. The generalization of the established principle to involve a certain amount of a priori information about the unknown parameter and further development of this new field of thermodynamics and quantum metrology would be definitely interesting and valuable in more realistic metrological scenarios. Besides, the generalized inequality [i.e. Eq.\,\eqref{Eq:entropy_QFI}] between the Shannon entropy under the SLD measurement and the QFI for both pure and mixed states replaces Eq.\,(3) in our Letter, and demonstrates a more clear relationship between the two information-theoretic quantities and deserves further investigation in order to gain more insights into these connections. 

\vspace{0.2cm}

We thank Shane Dooley and Martin B. Plenio for very helpful discussions.

\bibliography{reference}

\end{document}